\documentclass{article}

\usepackage{arxiv}

\usepackage[utf8]{inputenc} 
\usepackage[T1]{fontenc}    
\usepackage{hyperref}       
\usepackage{url}            
\usepackage{booktabs}       
\usepackage{amsfonts}       
\usepackage{nicefrac}       
\usepackage{microtype}      
\usepackage{lipsum}
\usepackage{graphicx}
\graphicspath{ {./images/} }
\usepackage{graphicx} 
\usepackage{tabularx}
\graphicspath{{./figures}}
\usepackage{longtable}
\usepackage{hyperref} 
\usepackage{float} 
\usepackage{pifont} 

\usepackage{array} 
\newcolumntype{L}[1]{>{\raggedright\arraybackslash}p{#1}} 
\newcolumntype{C}[1]{>{\centering\arraybackslash}p{#1}}   
\newcolumntype{R}[1]{>{\raggedleft\arraybackslash}p{#1}}  
\usepackage{url}
\usepackage[most]{tcolorbox}

\usepackage{booktabs}   
\usepackage{multirow}   

\title{MITRE-SAGE: A Multi-Agent Cybersecurity Question-Answering Model}

\author{
 Ali Habibzadeh \\
  Department of Computer Engineering\\
  University of Guilan\\
  Rasht, Iran \\
  \texttt{ali78@webmail.guilan.ac.ir} \\
   \And
 Farid Feyzi \\
  Department of Computer Engineering\\
  University of Guilan\\
  Rasht, Iran \\
  \texttt{feizi@guilan.ac.ir} \\
  \And
 Reza Ebrahimi Atani \\
  Department of Computer Engineering\\
  University of Guilan\\
  Rasht, Iran \\
  \texttt{rebrahimi@guilan.ac.ir} \\
}

\begin{document}
\maketitle
\begin{abstract}
Effective cybersecurity operations require timely and accurate analysis of large-scale heterogeneous security information; however, analysts increasingly struggle with information overload, alert fatigue, and time-constrained decision-making. Although large language models (LLMs) have demonstrated promising capabilities for question answering (QA), their effectiveness in cybersecurity remains limited by insufficient domain knowledge,  tendency to hallucinate, and difficulties in capturing both semantic and structural relationships.
This work proposes MITRE-SAGE, a multi-agent retrieval-augmented generation framework that integrates semantic and structural cybersecurity knowledge to improve the reliability and interpretability of LLM-based QA systems. By decomposing complex tasks into query interpretation, evidence retrieval, and answer synthesis, MITRE-SAGE effectively supports cybersecurity tasks such as vulnerability assessment, threat profiling, and relationship extraction. Furthermore, we proposed MITRE-QA, a comprehensive benchmark comprising 3,000 question–answer pairs for evaluating LLMs across diverse cybersecurity knowledge tasks, and used it to systematically evaluate MITRE-SAGE against representative baseline methods. Extensive experiments demonstrate that MITRE-SAGE consistently outperforms general-purpose and fine-tuned LLMs, as well as conventional RAG approaches. Notably, a lightweight configuration achieves superior performance on five of the eight benchmark tasks, indicating the effectiveness of the proposed multi-agent framework. The results highlight the potential of MITRE-SAGE as a scalable and interpretable approach for reliable cybersecurity QA, while MITRE-QA provides a standardized benchmark for future research.
The MITRE-QA benchmark is publicly available at \cite{habib2026mitreqa}.
\end{abstract}

\keywords{Cybersecurity \and Security Operation centers \and Generative AI \and Large language model \and Question Answering \and Chatbot \and Threat Intelligence}

\section{Introduction}\label{sec:intro}

In an era of rapid digital transformation, cybersecurity has become a critical component of global infrastructure, playing an essential role in ensuring business continuity and protecting organizational assets. However, securing increasingly complex digital environments is challenged by the massive volume, velocity, and heterogeneity of security data, which can overwhelm security personnel \cite{ali2025ai}. Security Operations Center (SOC) analysts operate in high-stakes environments where they must make timely and accurate decisions under severe time constraints while often having limited access to relevant and actionable information \cite{reeves2023understanding}. This challenge is further exacerbated by excessive false positives, resulting in alert fatigue, increased cognitive workload, and analyst burnout \cite{tariq2025alert, mohamed2025artificial}. Cisco's 2025 State of Security Report \cite{cisco2025state} found that 59\% of organizations struggle with excessive security alerts, 55\% report challenges caused by excessive false positives, and 57\% lose valuable investigation time due to fragmented security data, highlighting the growing operational burden on SOC analysts. Similarly, the ISC2 2025 Cybersecurity Workforce Study \cite{isc22025workforce} indicates that organizations continue to face significant cybersecurity workforce shortages, limiting their ability to effectively triage and investigate the continuously increasing volume of security events.

Traditional SOC infrastructures primarily rely on rule-based SIEM correlation and machine learning pattern classifiers, which are increasingly inadequate for interpreting the semantic complexity and contextual depth of heterogeneous security telemetry. These legacy systems often lack the cognitive reasoning capabilities required to provide actionable forensic insights and adapt to rapidly evolving attack patterns, underscoring the need for more sophisticated analytical paradigms \cite{gonzalez2021security, mohamed2025artificial, chopra2026chatnvd}. Recent advances in large language models (LLMs) have drawn significant attention to the development of question answering (QA) systems for cybersecurity, owing to their ability to automate security data analysis, extract actionable insights, and support analysts in reducing workload and improving decision-making efficiency \cite{vaswani2017attention, brown2020language}. Nevertheless, their performance diminishes significantly when deployed alone in mission-critical, domain-specific fields such as cybersecurity. Key contributing factors include over-reliance on static pre-training data, hallucination, lack of access to up-to-date information, and insufficient comprehension of highly specialized terminology \cite{huang2025survey, arefeen2024leancontext}. These limitations expose substantial risks in high-stakes environments where factual precision and logical consistency are essential.

Retrieval-Augmented Generation (RAG) has emerged as a prominent paradigm for QA, enabling LLMs to generate responses grounded in reliable, domain-specific knowledge and thereby improving the factual accuracy and verifiability of their outputs \cite{lewis2020retrieval}. Despite its widespread adoption, traditional RAG pipelines introduce several critical failure points. They exhibit high sensitivity to retrieval noise, where even a few irrelevant documents can mislead the generator \cite{zeng2025worse}. Additionally, most RAG implementations rely on flat vector representations that encode semantic similarity but fail to preserve relational structures, thereby undermining multi-hop reasoning \cite{guo2024lightrag}. Furthermore, conventional chunking strategies often fragment information, causing critical context to become misaligned across passages \cite{setty2024improving}.

\begin{figure}[H]
    \centering
    \includegraphics[width=0.8\linewidth]{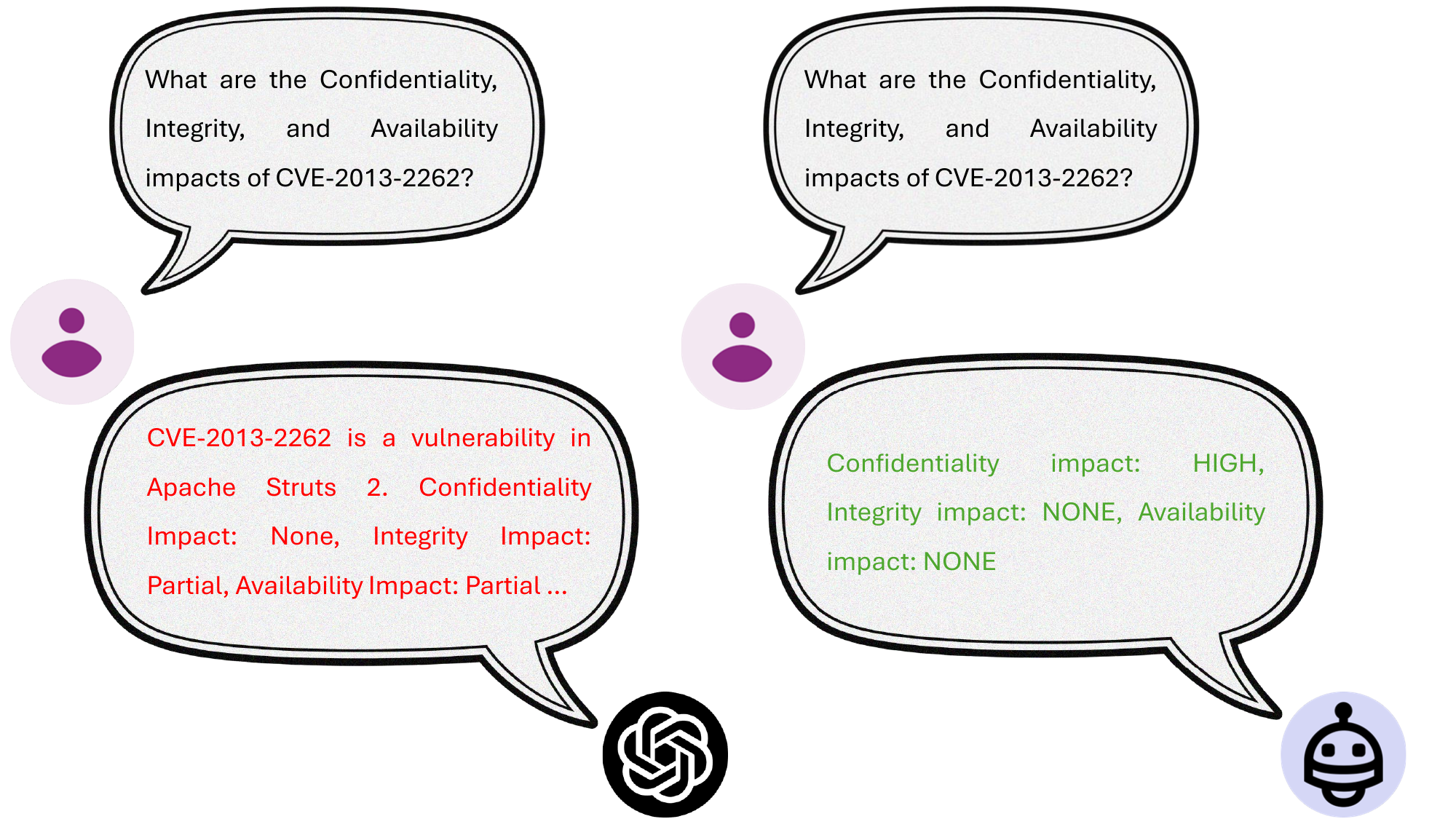}
    \caption{A comparative example of responses generated by GPT-4.1 and MITRE-SAGE.}
    \label{fig:example}
\end{figure}

To address these challenges, this paper proposes the MITRE-SAGE model (SAGE stands for Semantic Assistant for Guided Evidence), a multi-agent LLM-based framework that integrates knowledge graph, text, and web-based sources. MITRE-SAGE adopts a hierarchical multi-agent architecture consisting of an Orchestrator Agent, three specialized retrieval agents, and their corresponding Summarization Agents. The Orchestrator Agent coordinates the retrieval process and synthesizes the final response, while the retrieval agents acquire complementary evidence from heterogeneous knowledge sources. The Summarization Agents refine the retrieved information by filtering query-irrelevant content before forwarding the resulting evidence to the Orchestrator Agent for response generation. This collaborative framework provides three key advantages over existing approaches. First, leveraging graph-based representations enables a comprehensive understanding of domain concepts and inter-entity relationships. Second, it demonstrates strong performance not only on single-hop reasoning tasks but also on complex multi-hop reasoning scenarios. Third, by relying exclusively on valid and reliable information sources, the framework addresses concerns regarding the trustworthiness of language-based models among domain experts; an illustrative example of this process is shown in Figure \ref{fig:example}. Collectively, these capabilities assist cybersecurity professionals and organizations in making informed, evidence-based decisions for cyber threat analysis and response.

Furthermore, we developed MITRE-QA, a new cybersecurity benchmark designed to overcome the limitations of existing benchmarks and enable a more comprehensive and rigorous evaluation of model performance. The benchmark is primarily centered on the MITRE \cite{mitre_attack} and NVD \cite{nvd} knowledge bases, assessing both the semantic understanding and structural reasoning capabilities of models with respect to the entities and relationships represented within these resources. Additionally, MITRE-QA incorporates tasks designed to evaluate models' understanding of broader cybersecurity concepts beyond the scope of these knowledge bases. By integrating these complementary evaluation tasks, the proposed benchmark provides researchers with a holistic assessment of model performance across diverse aspects of the cybersecurity domain.

The main contributions of this paper are summarized as follows:

\begin{itemize}
	\item We propose a collaborative multi-agent RAG framework that enhances cybersecurity QA through trusted knowledge grounding, agent collaboration, and improved multi-hop reasoning capabilities.
	
	\item We introduce MITRE-QA, a comprehensive cybersecurity benchmark that evaluates LLMs on both semantic and structural knowledge understanding, including complex multi-hop reasoning tasks.
	
	\item We conduct extensive experiments against general-purpose and domain-specific LLMs, as well as hybrid RAG baselines, to demonstrate the effectiveness of the proposed framework.
	
\end{itemize}

The remainder of this paper is organized as follows. Section 2 reviews the related work, Section 3 introduces the proposed benchmark, Section 4 presents the proposed method, and Section 5 describes the experimental setup and evaluation results.

\section{Related Works}\label{sec:related}

A Question Answering (QA) model is a computational system designed to process natural language queries and retrieve precise information, representing an advanced paradigm of information retrieval. These systems have demonstrated transformative efficacy across diverse sectors, including medicine, finance, education, and law, where they streamline knowledge synthesis and support expert decision-making. Within the digital defense domain, specialized cybersecurity QA models act as AI-driven assistants that help security professionals investigate complex attacks, identify vulnerabilities, and attribute threats. These models significantly improve SOC efficiency by providing rapid access to curated threat intelligence and automating routine data sifting, which mitigates analyst alert fatigue and reduces the mean time to respond \cite{franco2020secbot}. Owing to the advancement of LLMs and their superior reasoning and generalization capabilities, these models have become a fundamental component of cybersecurity QA systems in research. Existing LLM-based cybersecurity QA frameworks are generally categorized into fine-tuning-based and RAG-based approaches \cite{habibzadeh2026large}.

\textbf{Fine-Tuning in Cybersecurity.}
Fine-tuning-based models involve the direct update of internal model parameters using specialized, domain-specific datasets to tailor the generator for specific security tasks \cite{paduraru2024cyberguardian, ji2024sevenllm, levi2025cyberpal}. These models are often impractical for implementation in operational cybersecurity environments due to their inability to adapt to the rapidly evolving threat landscape, as well as significant risks of catastrophic forgetting, hallucination, and the limited availability of high‑quality, expert‑curated training data.

\textbf{RAG in Cybersecurity.}
On the other hand, RAG-based frameworks integrate LLMs with external, up-to-date knowledge repositories, providing a mechanism to ground responses in verifiable domain information and mitigate hallucinations \cite{rajapaksha2024rag, chopra2026chatnvd, arikkat2024intellbot, zhang2026ambiguous, mitra2024localintel, simoni2025morse}. However, existing approaches remain limited by sensitivity to noisy documents, reliance on flat data representations that hinder the modeling of multi-hop semantic relationships, and conventional chunking strategies that can fragment critical contextual information. 

\textbf{LLM-based Agentic Systems.}
LLM-based agentic systems coordinate multiple specialized agents to solve complex tasks through structured interaction and functional role division. These autonomous frameworks enhance system capabilities by managing multi-step workflows and integrating external tools to handle intricate operations in diverse, domain-specific environments \cite{fang2025orion}. In this paper, we adopt a similar approach by designing a multi-agent framework that revolves around graphs, text, and the web. Graphs provide a highly suitable structural representation for cybersecurity knowledge, as they effectively capture high-level relationships and interactions among entities. Text retrieval complements graph-based reasoning by providing detailed contextual information, procedural descriptions, and semantic knowledge that may not be explicitly represented in structured graphs. Additionally, access to web-based knowledge sources is essential for incorporating the latest and domain-specific cybersecurity information. Moreover, given the critical and high-stakes nature of cybersecurity decisions, the integration of verifiable evidences within model-generated answers offers a promising pathway to mitigate fundamental challenges inherent to LLMs, including hallucination, bias, inconsistency, and opacity. Such grounding not only addresses these vulnerabilities but also helps restore trust and confidence among cybersecurity practitioners \cite{habibzadeh2026large}.

\section{MITRE-QA Benchmark}\label{sec:benhmark}
MITRE-QA is a cybersecurity benchmark comprising 3,000 question-answer pairs constructed from reliable cybersecurity knowledge sources, including MITRE and the NVD. The benchmark is designed to evaluate the capability of language models in understanding and reasoning over cybersecurity knowledge.

\subsection{Benchmark Construction Methodology}
The proposed benchmark comprises eight tasks, six of which were newly designed from scratch based on authoritative cybersecurity knowledge sources, including the MITRE ATT\&CK, CAPEC, CWE, and NVD CVE datasets. The remaining two tasks were adapted from existing benchmarks and categorized as out-of-distribution (OOD) tasks, as they are not directly represented within the knowledge sources utilized by MITRE-SAGE. 

For five out of the six tasks, questions were constructed through a template-based question generation pipeline implemented in Python, where manually designed templates were instantiated using entities, attributes, and relationships extracted from the underlying cybersecurity datasets. For the Entity Identification task, instead of using predefined templates, the relevant document were provided to GPT-4.1, which was instructed to generate corresponding questions based on the provided content. The generated questions support both descriptive and multiple-choice answer formats. Multiple-choice answers were manually annotated, whereas descriptive answers were generated using GPT-4.1 by providing the model with each question and its corresponding supporting documents. Finally, all question–answer pairs underwent manual validation by the second and third authors of this study, who possess expertise in both cybersecurity research and educational assessment, to ensure correctness, relevance, and clarity.

\subsection{Task Descriptions}
To ensure that the proposed benchmark is comprehensive and systematically evaluates all relevant aspects of a QA model in the cybersecurity domain, we categorize the tasks into two types of difficulty: single-hop reasoning and multi-hop reasoning. Single-hop reasoning tasks are those that require retrieving information from a single knowledge source or traversing one retrieval agent to derive the answer, while multi-hop reasoning tasks are those that necessitate the integration of information across multiple knowledge sources, demanding more complex inference and compositional understanding.

\subsubsection{Single-hop}
At this level, each question can be resolved through a single interaction between the Orchestrator Agent and one of its subordinate agents. 
The tasks defined at this level are as follows:
\begin{itemize}
	\item \textbf{Conceptual Understanding}: This task involves answering direct, definition-based, and explanatory questions about cybersecurity concepts and standards. It requires entity recognition, terminology disambiguation, and reliable knowledge grounding to prevent hallucination. We extracted question answer pairs from the dataset introduced in \cite{zhao2025ontology}.
	
	\item \textbf{Entity Attribute Retrieval}: This task refers to extracting and presenting authoritative, organized information about explicitly specified cybersecurity entities from standardized knowledge bases. In this task, the model is queried regarding the intrinsic characteristics, attributes, or metadata of a given entity. This task encompasses text summarization, attribute extraction, and entity-centric factual question answering.
	
	\item \textbf{Structured Knowledge Retrieval}: This task requires a model to answer questions that involve traversing relationships within a structured cybersecurity knowledge graph. Instead of returning isolated facts, this task focuses on schema-aware reasoning over entities and their relations. Effective performance on this task requires a model to accurately capture relational dependencies and reason over interconnected entity structures within the knowledge graph.

	\item \textbf{Entity Identification}: 
	 This task requires a model to map a natural language description of a cybersecurity entity to its corresponding standardized identifier within established taxonomies. In this setting, the model receives a textual characterization, potentially including behavioral patterns, technical properties, or contextual attributes, and must determine the most appropriate entity label. This task assesses the model’s ability to perform fine-grained semantic matching and accurate entity grounding in cybersecurity knowledge bases.
	
\end{itemize}

\subsubsection{Multi-hop}
Questions at this level necessitate a minimum of two interactions between the Orchestrator Agent and one of its subordinate agents. Similar to the previous category, responses may be either descriptive or multiple-choice in nature.
The tasks defined at this level are as follows:

\begin{itemize}
	\item \textbf{Relation-Aware Entity Attribute Retrieval}: This is the task of identifying and synthesizing attributes of a specified cybersecurity entity that are defined through its relationships with other entities within a structured threat intelligence framework. In this setting, the model is prompted to retrieve and explain properties that emerge from inter-entity associations such as the sectors targeted by a threat actor, the campaigns attributed to a specific group, or the techniques employed within a particular operation. Rather than focusing solely on intrinsic attributes, this task emphasizes relationship-aware reasoning, requiring traversal of structured links and aggregation of distributed evidence. Successful execution necessitates accurate entity recognition, multi-hop relational reasoning, contextual synthesis, and faithful grounding in authoritative threat intelligence sources to ensure analytical reliability and semantic precision.
	
	\item \textbf{Threat Profiling}: This task focuses on reasoning over threat actors and malicious software artifacts. The assistant must support multi-step inference, such as identifying techniques or tools commonly associated with a given threat group or malware family. It requires relational reasoning, aggregation across reports, and contextual interpretation of threat intelligence data.
	
	\item \textbf{Relation-Aware Entity Identification}: This task requires a model to identify a cybersecurity entity based on user-provided inter-entity relationships and descriptive criteria. This task combines relational reasoning with description-driven entity identification, enabling the system to first enumerate connected entities such as associated tactics, related campaigns, or dependent software, and then locate the target entity based on the provided property. Effective execution requires precise schema-aware traversal of relationships, semantic matching, and aggregation of contextual information, ensuring that all retrieved linked entities are relevant and accurately reflect the underlying structured cybersecurity knowledge.
	
	\item \textbf{Sigma Rules to Attack Techniques Mapping}: This task focuses on automatically identifying the corresponding MITRE ATT\&CK techniques associated with a given Sigma detection rule. In this task, a model is provided with the semantic components of a Sigma rule, including its title, description, log source, and detection logic, and is required to infer the most relevant ATT\&CK techniques that describe the underlying adversarial behavior. It evaluates the capability of cybersecurity intelligence systems to perform semantic reasoning over detection rules and to understand the relationship between low-level detection patterns and high-level adversarial techniques defined in the MITRE ATT\&CK framework. The question answer pairs obtained from SigmaHQ \cite{sigmahq_sigma}.

\end{itemize}

\begin{table*}[t]
	\centering
	\caption{Overview of the tasks included in the MITRE-QA benchmark.}
	\label{tab:realted_analysis}
	
	\small
	\begin{tabularx}{\textwidth}{p{3cm} p{2.5cm} p{2cm} p{2cm} X}
		\hline
		\textbf{Task} & \textbf{Reasoning complexity} & \textbf{Answer format} & 
		\textbf{Samples} & \textbf{Metadata} \\
		\hline
		
		Conceptual Understanding & Single-hop & Text Generation & 500 & OOD task and extracted from \cite{zhao2025ontology}. \\ \hline
		
		Entity Attribute Retrieval & Single-hop & Text Generation & 500 & The entity distribution for this task is CVE (70\%), CWE (10\%), CAPEC (10\%), and Other (10\%). \\ \hline
		
		Structured Knowledge Retrieval & Single-hop & Multiple Choice & 500 & The hop distribution for this task is 1-hop (70\%), 2-hop (20\%), and 3-hop (10\%). \\ \hline
		
		Entity Identification & Single-hop & Multiple Choice & 500 & The entity distribution for this task is CVE (70\%), CWE (10\%), CAPEC (10\%), and Other (10\%). \\ \hline
		
		Relation-Aware Entity Attribute Retrieval & Multi-hop & Text Generation & 250 & The entity distribution is balanced across all entity types. \\ \hline
		
		Threat Profiling & Multi-hop & Text Generation & 250 & Only Malware and Group entities are used for this task. \\ \hline
		Relation-Aware Entity Identification & Multi-hop & Multiple Choice & 250 & Only CVE, CWE, and CAPEC entities are used for this task. \\ \hline
		
		Sigma Rules to Attack Technique Mapping & Multi-hop & Multiple Choice & 250 & OOD task and extracted from \cite{sigmahq_sigma}. \\ \hline
		
		\hline
	\end{tabularx}
\end{table*}

\section{Method}\label{sec:model}

\begin{figure}[H]
    \centering
    \includegraphics[width=0.8\linewidth]{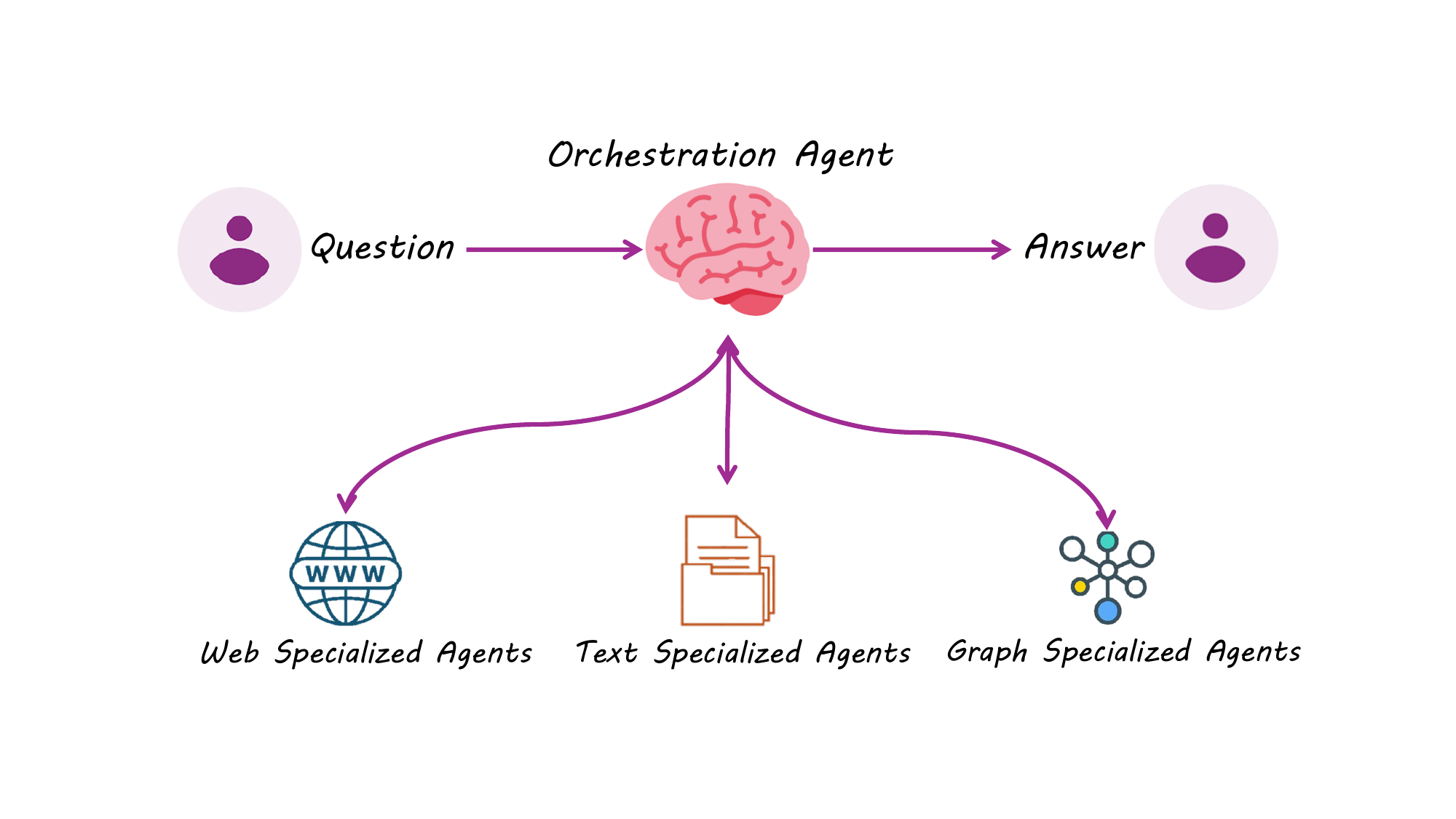}
    \caption{Architecture of the proposed MITRE-SAGE framework.}
    \label{fig:method}
\end{figure}

As illustrated in Figure \ref{fig:method}, MITRE-SAGE consists of an Orchestrator Agent and three specialized retrieval-based agent groups, each designed to access and process information from a distinct knowledge source. Each agent group comprises a Retrieval Agent and a corresponding Summarization Agent. The Orchestrator Agent serves as the central controller, responsible for analyzing the user query, determining the required retrieval sources, coordinating agent interactions, and synthesizing the collected evidence to generate the final response. The Retrieval Agents focus on acquiring relevant information from their dedicated knowledge sources, while the Summarization Agents refine the retrieved information by eliminating irrelevant content, reducing redundancy, and extracting query-relevant evidence before forwarding it to the Orchestrator Agent.

\subsection{Orchestrator Agent}\label{sec:o-agent}
The Orchestrator Agent is responsible for determining the required agents and generating appropriate sub-queries to delegate each agent based on the user's request. The Orchestrator is guided by chain-of-thought prompting with few-shot examples, ensuring
interpretable step-wise decomposition that supports grounded reasoning in downstream modules. After receiving evidence from the selected agents, the Orchestrator Agent evaluates the retrieved information and determines whether additional agent consultations are required to obtain sufficient evidence for answering the query. Finally, it synthesizes the available evidence and generates the final response provided to the user.

\subsection{Graph Retrieval Agent}\label{sec:g-agent}
\begin{figure}[H]
    \centering
    \includegraphics[width=0.8\linewidth]{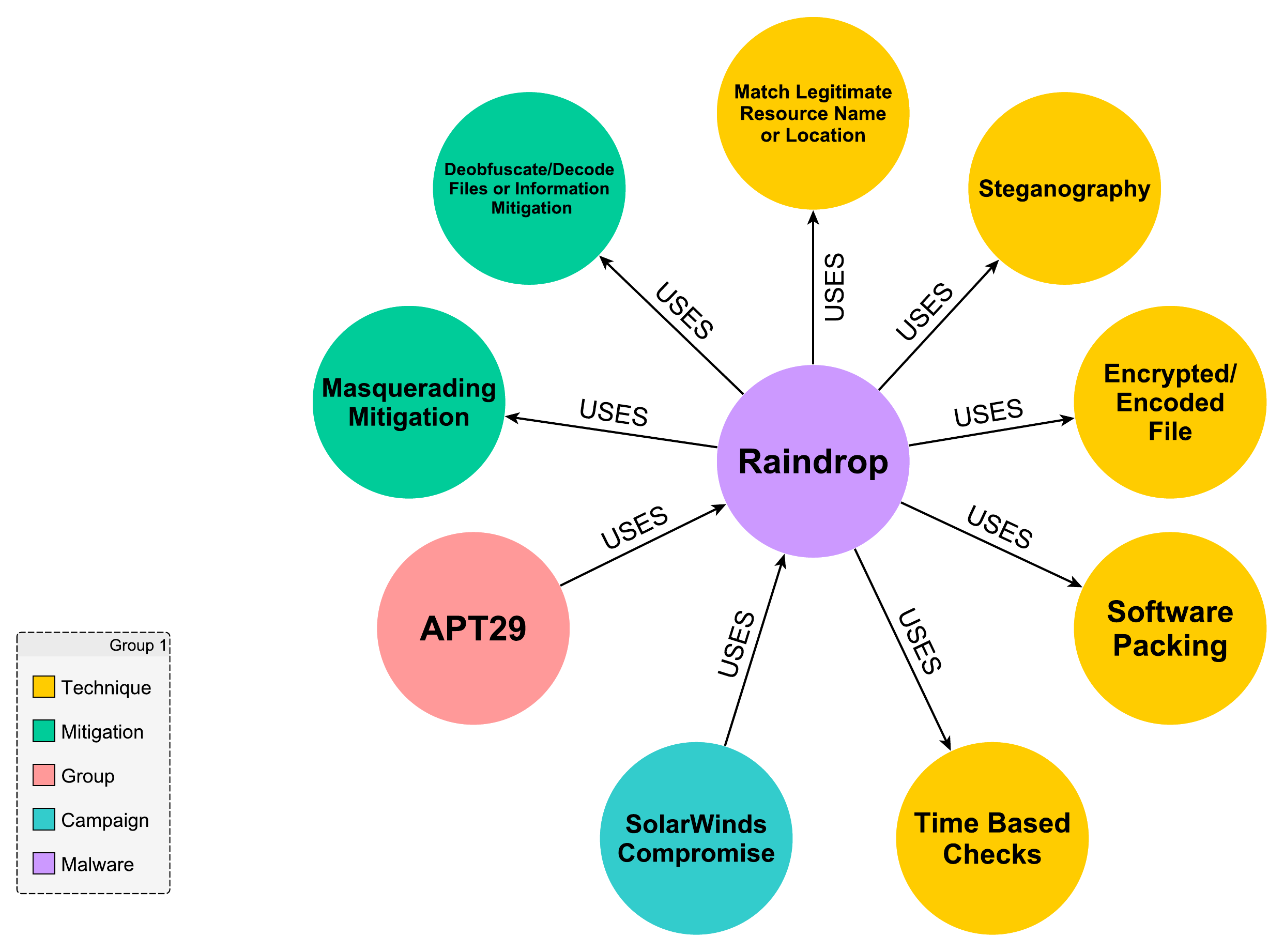}
    \caption{Subset of the cybersecurity knowledge graph constructed from MITRE and NVD.}
    \label{fig:subgraph}
\end{figure}

This agent is responsible for generating Cypher queries corresponding to the queries received from the Orchestrator Agent.  Then, the generated Cypher query is executed on a cybersecurity knowledge graph containing entities and relationships extracted from the MITRE and NVD knowledge bases. A subset of this knowledge graph is illustrated in Figure \ref{fig:subgraph}.

To enhance the agent's capability in producing accurate and domain-specific Cypher queries, the agent was optimized through soft prompt tuning on the text-to-Cypher generation task. In this approach, a sequence of trainable virtual tokens $P_v$ is introduced while keeping the original language model parameters frozen. The optimized input representation can be formulated as:

\begin{equation}
	X' = [P_v; X_q]
\end{equation}

where $P_v = \{p_1,p_2,...,p_m\}$ denotes the set of learnable virtual tokens and $X_q$ represents the tokenized input query. During training, only the parameters of the soft prompt are updated by minimizing the Cypher generation loss:

\begin{equation}
	\mathcal{L} = -\sum_{t=1}^{T} \log P(c_t|c_{<t},X')
\end{equation}

where $c_t$ denotes the $t$-th token of the target Cypher query and $T$ represents the length of the generated query sequence.

\subsection{Text Retrieval Agent}\label{sec:t-agent}
This agent is responsible for retrieving relevant information from textual documents based on the query provided by the Orchestrator Agent. Depending on the nature of the query, it employs one of two retrieval mechanisms: a sparse retriever, which is optimized for identifying entity identifiers and extracting their associated attributes, and a hybrid retriever, which is designed to identify the corresponding entities.

The sparse retriever performs lexical retrieval by ranking documents according to their keyword relevance to the generated sub-query. To leverage both lexical matching and semantic similarity, the hybrid retriever combines the rankings produced by the sparse and dense retrievers using the Reciprocal Rank Fusion (RRF) algorithm. Given a document $d$, its final retrieval score is computed as:

\begin{equation}
	\mathrm{RRF}(d)=\sum_{i=1}^{N}\frac{1}{k+\mathrm{rank}_i(d)},
\end{equation}

where $N$ denotes the number of retrieval methods, $\mathrm{rank}_i(d)$ is the rank assigned to document $d$ by the $i$-th retriever, and $k$ is a constant controlling the influence of individual rankings. Documents are subsequently ranked according to their RRF scores, and the highest-ranked documents are returned as the final retrieval results.

\subsection{Web Retrieval Agent}\label{sec:w-agent}
The primary objective of this agent is to retrieve up-to-date information related to a specific topic and assist the Orchestrator Agent in interpreting technical and domain-specific terms. Furthermore, this agent serves as a complementary information source when the other retrieval agents are unable to provide sufficient evidence for answering the user query.

\subsection{Summarization Agent}\label{sec:s-agent}
Each retrieval agent group incorporates a dedicated Summarization Agent responsible for processing the raw retrieval results obtained from its corresponding Retrieval Agent. The Summarization Agent analyzes the retrieved information, extracts query-relevant evidence, removes irrelevant or redundant content, and forwards the refined information to the Orchestrator Agent. This strategy preserves reasoning continuity while maintaining a concise context representation, ultimately supporting more accurate and efficient answer generation.

\section{Experiments}\label{sec:result}

\subsection{Baselines and Evaluation Metrics}

To evaluate the effectiveness of the proposed framework, we compare it against three types of baselines. First, we select GPT-4.1, Qwen2.5-32B, Gemma-3-27b-it as strong general-purpose baselines due to the advanced reasoning capabilities across diverse tasks, including cybersecurity applications \cite{alam2025athenabench, liu2024cyberbench}. Second, we adopt Foundation-Sec-8B-Instruct as a representative cybersecurity-specialized model, which is based on Llama 3.1 8B, continued pre-trained on approximately 5.1 billion tokens of cybersecurity-specific data, and subsequently fine-tuned for instruction following and conversational cybersecurity tasks \cite{weerawardhena2025llama}. Third, we develop a conventional hybrid RAG-based approach using Qwen2.5-32B as the generation model, combining dense and sparse retrieval mechanisms. Unlike conventional document chunking strategies that divide text based on fixed character lengths, we represent each cybersecurity entity and its associated attributes as an individual document. This design preserves the semantic integrity and completeness of entity-level knowledge, which is essential for accurate cybersecurity information retrieval.

For multiple-choice question answering tasks, we adopt accuracy as the primary evaluation metric, calculated as the proportion of correctly answered questions. For text generation tasks, we employ the RAGAS evaluation framework \cite{es2024ragas} and measure three complementary dimensions: answer relevance, answer similarity, and answer correctness. The evaluation is performed using GPT-4.1-mini as the LLM-based evaluator and text-embedding-3-large as the embedding model.

\textbf{Answer Relevance.}
Answer relevance measures how well the generated response addresses the original query. Following RAGAS, synthetic questions $\{q_1, q_2, ... , q_n\}$ are generated from the generated answer, and the relevance score is computed as the average cosine similarity between their embeddings and the original query embedding:

\[
R_{\mathrm{rel}}(q,a_g)=\frac{1}{n}\sum_{i=1}^{n}
\frac{E(q_i)\cdot E(q)}
{\|E(q_i)\|\|E(q)\|}
\]

where $E(\cdot)$ denotes the embedding function, $q$ is the original query, and $q_i$ represents the generated questions. In this study, we follow the default configuration of RAGAS and set $n=3$, meaning that three synthetic questions are generated for each response to estimate the relevance score.

\textbf{Answer Similarity.}
Answer similarity evaluates the semantic consistency between the generated answer and the reference answer. Given a generated answer $a_g$ and a reference answer $a_r$, the similarity score is computed using the cosine similarity between their embedding representations:
\[
R_{\mathrm{sim}}(a_g,a_r)=
\frac{E(a_g)\cdot E(a_r)}
{\|E(a_g)\|\|E(a_r)\|}
\]
where $E(\cdot)$ denotes the embedding function provided by the text-embedding-3-large model.

\textbf{Answer Correctness.}
Answer correctness evaluates the factual consistency of the generated response with respect to the reference answer. Following the RAGAS framework, the correctness score combines an LLM-based factual evaluation with semantic similarity between the generated and reference answers:

\[
R_{\mathrm{corr}}(a_g,a_r)
=
\alpha \, R_{\mathrm{fact}}(a_g,a_r)
+
(1-\alpha)\,R_{\mathrm{sim}}(a_g,a_r),
\]

where $a_g$ and $a_r$ denote the generated and reference answers, respectively, $R_{\mathrm{fact}}(\cdot)$ is the factual correctness score assigned by the LLM evaluator, $R_{\mathrm{sim}}(\cdot)$ is the semantic similarity score computed from the answer embeddings, and $\alpha$ is the weighting coefficient, which is set to $0.75$.

\subsection{Implementation Details}

All experiments conducted in this study were performed on a single NVIDIA A100 GPU with 80 GB of memory. The proposed framework is implemented using Qwen2.5-14B as the Orchestrator Agent and Qwen2.5-7B as the underlying model for all subordinate agents. Each model is deployed with 8-bit quantization and generates responses with a maximum length of 1,024 tokens. The Orchestrator and retrieval agents employed a few-shot prompting strategy to improve task understanding and guide the generation of appropriate responses. The Orchestrator was constrained to invoke each retrieval agent at most once during a single reasoning process, thereby preventing redundant retrieval operations. All adopted prompt templates used in this study are provided in \cite{habib2026mitreqa}.

The Graph Retrieval Agent utilizes an Aura Neo4j server for executing Cypher queries and retrieving structured cybersecurity knowledge. To enhance Text2Cypher generation, we perform soft prompt tuning following the prompt design strategy proposed in \cite{mandilara2025decoding}. The model trained using 10,000 training samples with 80 virtual tokens for five epochs. We employ a learning rate of 2e-2 and warmup ratio of 0.03.

For the hybrid retriever, the dense and sparse retrieval components independently retrieve the top-50 candidate documents. Subsequently, RRF is employed to combine and re-rank the retrieved results, from which the top-10 documents are selected as the final retrieval set. These values are doubled for the Hybrid RAG baseline to leverage the larger context window and stronger retrieval capabilities of the more powerful LLM employed in that baseline. The sparse retriever utilizes the BM25 algorithm, while the dense retriever employs SecureBERT2.0-biencoder for generating cybersecurity-specific embeddings \cite{aghaei2025securebert}.

The Web Retrieval Agent employs the DuckDuckGo search API to retrieve complementary external information when required. Given a orchestrator query, the agent is instructed to generate three search queries and retrieve the top three results for each query to improve coverage and increase the likelihood of obtaining relevant information. DuckDuckGo is selected due to its free accessibility and lightweight integration, enabling cost-effective retrieval of up-to-date cybersecurity information.

\subsection{Results}\label{sec:eval}

Tables \ref{tab:single-hop-qa} and \ref{tab:multi-hop-qa} compare the performance of MITRE-SAGE with the baseline methods on the MITRE-QA benchmark. MITRE-SAGE achieves the best overall performance across five of the eight benchmark tasks while demonstrating competitive performance on the remaining three tasks. Moreover, compared with the strongest baseline, it achieves improvements of up to 58\% in multiple-choice accuracy and up to 35.5\% in answer correctness for text generation tasks.
Moreover, its consistent performance across both single-hop and multi-hop reasoning tasks indicates that the proposed framework generalizes effectively to cybersecurity questions with varying levels of reasoning complexity.

The performance of baseline methods declines substantially on Text Generation of multi-hop reasoning tasks compared with single-hop tasks, indicating limited capability to effectively integrate and reason over multiple pieces of interconnected cybersecurity knowledge. General purpose models achieves the best performance on conceptual understanding and sigma rules to attack technique mapping tasks, which is consistent with its strong general-purpose language understanding capabilities. On other hand, the hybrid RAG baseline demonstrate the best performance on the Entity Identification task, reflecting the effectiveness of retrieval-based approaches for matching descriptive information to specific cybersecurity entities. Moreover, Foundation-Sec-8B-Instruct exhibited the weakest performance, which indicate the limitations of an 8-billion-parameter language model in retaining and inferring complex cybersecurity concepts.

Overall, our proposed framework effectively combines the strong general reasoning capabilities of standalone LLMs with the domain-specific knowledge grounding of retrieval-augmented approaches, enabling accurate responses to both general-purpose and expert-level cybersecurity questions. These results demonstrate that the proposed collaborative architecture effectively bridges the gap between the reasoning capabilities of general-purpose LLMs and the reliability required for trustworthy cybersecurity question answering.

\begin{table}[htbp]
	\centering
	\caption{Comparative results on single-hop tasks of MITRE-QA. The evaluated tasks include Structured Knowledge Retrieval (SKR), Entity Identification (EI), Conceptual Understanding (CU), and Entity Attribute Retrieval (EAR).}
	\label{tab:single-hop-qa}
	\begin{tabular}{lcc|ccc|ccc}
		\toprule
		\multirow{3}{*}{Model} &
		\multicolumn{2}{c|}{Multiple Choice} &
		\multicolumn{6}{c}{Text Generation} \\
		\cmidrule(lr){2-3} \cmidrule(lr){4-9}
		& SKR & EI &
		\multicolumn{3}{c|}{CU} &
		\multicolumn{3}{c}{EAR} \\
		\cmidrule(lr){2-2} \cmidrule(lr){3-3}
		\cmidrule(lr){4-6} \cmidrule(lr){7-9}
		& Acc. & Acc. & Rel. & Sim. & Corr. & Rel. & Sim. & Corr. \\
		\midrule
		GPT-4.1 & \underline{0.310} & 0.620 & \textbf{0.922} & \underline{0.830} & 0.519 & \underline{0.756} & \underline{0.674} & \underline{0.353} \\
        Qwen2.5-32B & 0.282 & 0.360 & \textbf{0.922} & \textbf{0.849} & \textbf{0.577} & 0.527 & 0.656 & 0.310 \\
        Gemma-3-27b-it & 0.268 & 0.346 & 0.883 & 0.806 & \underline{0.521} & \textbf{0.770} & 0.642 & 0.283 \\
        Foundation-Sec-8B-Instruct & 0.260 & 0.300 & \underline{0.902} & 0.800 & 0.405 & 0.478 & 0.634 & 0.272 \\
		Hybrid RAG & 0.228 & \textbf{0.947} & 0.856 & 0.812 & 0.456 & 0.175 & 0.649 & 0.331 \\
		MITRE-SAGE & \textbf{0.890} & \underline{0.852} & 0.829 & 0.802 & 0.472 & 0.677 & \textbf{0.788} & \textbf{0.532} \\
		\bottomrule
	\end{tabular}
\end{table}

\begin{table}[htbp]
	\centering
	\caption{Comparative results on multi-hop tasks of MITRE-QA. The evaluated tasks include Relation-Aware Entity Identification (REI), Sigma Rules to Attack Techniques Mapping (SR2AT), Relation-Aware Entity Attribute Retrieval (REAR), and Threat Profiling (TP).}
	\label{tab:multi-hop-qa}
	\begin{tabular}{lcc|ccc|ccc}
		\toprule
		\multirow{3}{*}{Model} &
		\multicolumn{2}{c|}{Multiple Choice} &
		\multicolumn{6}{c}{Text Generation} \\
		\cmidrule(lr){2-3} \cmidrule(lr){4-9}
		& REI & SR2AT &
		\multicolumn{3}{c|}{REAR} &
		\multicolumn{3}{c}{TP} \\
		\cmidrule(lr){2-2} \cmidrule(lr){3-3}
		\cmidrule(lr){4-6} \cmidrule(lr){7-9}
		& Acc. & Acc. & Rel. & Sim. & Corr. & Rel. & Sim. & Corr. \\
		\midrule
		GPT-4.1    & \underline{0.384} & \textbf{0.860} & \textbf{0.628} & \underline{0.408} & \underline{0.162} & \textbf{0.520} & \underline{0.543} & \underline{0.156} \\
        Qwen2.5-32B & 0.272 & 0.416 & 0.451 & 0.397 & 0.139 & 0.098 & 0.398 & 0.103 \\
        Gemma-3-27b-it & 0.248 & 0.368 & \underline{0.568} & 0.407 & 0.136 & \underline{0.466} & 0.479 & 0.122 \\
        Foundation-Sec-8B-Instruct & 0.048 & 0.352 & 0.438 & \underline{0.408} & 0.127 & 0.389 & 0.512 & 0.135 \\
		Hybrid RAG & 0.324 & 0.540 & 0.023 & 0.350 & 0.141 & 0.009 & 0.345 & 0.128 \\
		MITRE-SAGE & \textbf{0.724} & \underline{0.652} & 0.499 & \textbf{0.652} & \textbf{0.488} & 0.380 & \textbf{0.785} & \textbf{0.511} \\
		\bottomrule
	\end{tabular}
\end{table}

\section{Conclusion}\label{sec:conclusion}

In this paper, we introduced MITRE-SAGE, a novel hierarchical multi-agent RAG framework designed to address the limitations of existing LLM-based cybersecurity QA approaches. By integrating both structural and semantic representations of cybersecurity knowledge, MITRE-SAGE effectively captures high-level relationships among cybersecurity entities while retrieving fine-grained textual information associated with them. Furthermore, the integration of web-based retrieval as a complementary source enables the framework to incorporate general cybersecurity concepts and up-to-date information when required.
The proposed framework provides several advantages, particularly for resource-constrained security organizations. First, it reduces the dependency on costly and time-consuming model retraining or fine-tuning when new cybersecurity knowledge emerges, while mitigating the limitations of conventional RAG systems, including sensitivity to noisy retrieval results and reliance on a single knowledge modality. Second, MITRE-SAGE achieves effective cybersecurity assistance using resource-efficient open-source models, reducing computational requirements and addressing concerns associated with the adoption of third-party proprietary models. Finally, by providing automated knowledge retrieval, reasoning, and synthesis capabilities, the proposed framework reduces the expertise burden on security personnel and saves valuable time required for information gathering and analysis, enabling more efficient cybersecurity decision-making.  
Furthermore, we introduced MITRE-QA, a comprehensive benchmark comprising both single-hop and multi-hop reasoning tasks, and evaluated MITRE-SAGE against baseline approaches on this benchmark. The experimental results demonstrate that MITRE-SAGE achieves superior performance on five out of eight tasks, with improvements of up to 58\% on structured knowledge retrieval task. We expect further performance gains with the adoption of more capable language models and increased computational resources, without requiring fundamental modifications to the overall architecture. As a future research direction, the knowledge sources can be expanded by incorporating cyber threat intelligence reports and other cybersecurity corpora into the text vector store, as well as by extracting entities and relationships from these resources to enrich the cybersecurity knowledge graph, thereby improving knowledge coverage and reasoning capabilities. Furthermore, exploring unified models that combine unstructured text and structured graph data through advanced graph reasoning and tight-coupling GraphRAG architectures remains a promising area for research to enable deeper insights across heterogeneous knowledge ecosystems.

\bibliographystyle{unsrt}  
\bibliography{references}

@misc{habib2026mitreqa,
  author       = {Ali Habibzadeh},
  title        = {MITRE-QA: A Cybersecurity Question Answering Benchmark},
  year         = {2026},
  howpublished = {\url{https://github.com/alihabib9999/MITRE-QA}},
  note         = {GitHub repository},
}

@article{ali2025ai,
  title={AI-driven fusion with cybersecurity: Exploring current trends, advanced techniques, future directions, and policy implications for evolving paradigms--A comprehensive review},
  author={Ali, Sijjad and Wang, Jia and Leung, Victor Chung Ming},
  journal={Information Fusion},
  volume={118},
  pages={102922},
  year={2025},
  publisher={Elsevier}
}

@article{reeves2023understanding,
  title={Understanding decision making in security operations centres: building the case for cyber deception technology},
  author={Reeves, Andrew and Ashenden, Debi},
  journal={Frontiers in Psychology},
  volume={14},
  pages={1165705},
  year={2023},
  publisher={Frontiers Media SA}
}

@article{tariq2025alert,
  title={Alert fatigue in security operations centres: Research challenges and opportunities},
  author={Tariq, Shahroz and Baruwal Chhetri, Mohan and Nepal, Surya and Paris, Cecile},
  journal={ACM Computing Surveys},
  volume={57},
  number={9},
  pages={1--38},
  year={2025},
  publisher={ACM New York, NY}
}

@article{mohamed2025artificial,
  title={Artificial intelligence and machine learning in cybersecurity: a deep dive into state-of-the-art techniques and future paradigms},
  author={Mohamed, Nachaat},
  journal={Knowledge and Information Systems},
  volume={67},
  number={8},
  pages={6969--7055},
  year={2025},
  publisher={Springer}
}

@techreport{cisco2025state,
  author       = {{Cisco}},
  title        = {2025 State of Security Report},
  institution  = {Cisco Systems},
  year         = {2025},
  url          = {https://newsroom.cisco.com/c/r/newsroom/en/us/a/y2025/m05/global-state-of-security-report-reveals-critical-need-for-connected-security-operations.html},
  note         = {Accessed: 2026-07-25}
}

@techreport{isc22025workforce,
  author       = {{ISC2}},
  title        = {2025 Cybersecurity Workforce Study},
  institution  = {ISC2},
  year         = {2025},
  url          = {https://www.isc2.org/Insights/2025/12/2025-ISC2-Cybersecurity-Workforce-Study},
  note         = {Accessed: 2026-07-25}
}

@article{gonzalez2021security,
  title={Security information and event management (SIEM): analysis, trends, and usage in critical infrastructures},
  author={Gonz{\'a}lez-Granadillo, Gustavo and Gonz{\'a}lez-Zarzosa, Susana and Diaz, Rodrigo},
  journal={Sensors},
  volume={21},
  number={14},
  pages={4759},
  year={2021},
  publisher={MDPI}
}

@article{chopra2026chatnvd,
  title={Chatnvd: Advancing cybersecurity vulnerability assessment with large language models},
  author={Chopra, Shivansh and Ahmad, Hussain and Goel, Diksha and Szabo, Claudia},
  journal={IEEE Access},
  year={2026},
  publisher={IEEE}
}

@article{vaswani2017attention,
  title={Attention is all you need},
  author={Vaswani, Ashish and Shazeer, Noam and Parmar, Niki and Uszkoreit, Jakob and Jones, Llion and Gomez, Aidan N and Kaiser, {\L}ukasz and Polosukhin, Illia},
  journal={Advances in neural information processing systems},
  volume={30},
  year={2017}
}

@article{brown2020language,
  title={Language models are few-shot learners},
  author={Brown, Tom and Mann, Benjamin and Ryder, Nick and Subbiah, Melanie and Kaplan, Jared D and Dhariwal, Prafulla and Neelakantan, Arvind and Shyam, Pranav and Sastry, Girish and Askell, Amanda and others},
  journal={Advances in neural information processing systems},
  volume={33},
  pages={1877--1901},
  year={2020}
}

@article{huang2025survey,
  title={A survey on hallucination in large language models: Principles, taxonomy, challenges, and open questions},
  author={Huang, Lei and Yu, Weijiang and Ma, Weitao and Zhong, Weihong and Feng, Zhangyin and Wang, Haotian and Chen, Qianglong and Peng, Weihua and Feng, Xiaocheng and Qin, Bing and others},
  journal={ACM Transactions on Information Systems},
  volume={43},
  number={2},
  pages={1--55},
  year={2025},
  publisher={ACM New York, NY}
}

@article{arefeen2024leancontext,
  title={Leancontext: Cost-efficient domain-specific question answering using llms},
  author={Arefeen, Md Adnan and Debnath, Biplob and Chakradhar, Srimat},
  journal={Natural Language Processing Journal},
  volume={7},
  pages={100065},
  year={2024},
  publisher={Elsevier}
}

@article{lewis2020retrieval,
  title={Retrieval-augmented generation for knowledge-intensive nlp tasks},
  author={Lewis, Patrick and Perez, Ethan and Piktus, Aleksandra and Petroni, Fabio and Karpukhin, Vladimir and Goyal, Naman and K{\"u}ttler, Heinrich and Lewis, Mike and Yih, Wen-tau and Rockt{\"a}schel, Tim and others},
  journal={Advances in neural information processing systems},
  volume={33},
  pages={9459--9474},
  year={2020}
}

@article{zeng2025worse,
  title={Worse than zero-shot? a fact-checking dataset for evaluating the robustness of rag against misleading retrievals},
  author={Zeng, Linda and Gupta, Rithwik and Motwani, Divij and Zhang, Yi and Yang, Diji},
  journal={arXiv preprint arXiv:2502.16101},
  year={2025}
}

@article{guo2024lightrag,
  title={Lightrag: Simple and fast retrieval-augmented generation},
  author={Guo, Zirui and Xia, Lianghao and Yu, Yanhua and Ao, Tian and Huang, Chao},
  journal={arXiv preprint arXiv:2410.05779},
  volume={2},
  number={3},
  year={2024}
}

@article{setty2024improving,
  title={Improving retrieval for rag based question answering models on financial documents},
  author={Setty, Spurthi and Thakkar, Harsh and Lee, Alyssa and Chung, Eden and Vidra, Natan},
  journal={arXiv preprint arXiv:2404.07221},
  year={2024}
}

@misc{mitre_attack,
  author       = {{MITRE}},
  title        = {MITRE ATT\&CK: Adversarial Tactics, Techniques, and Common Knowledge},
  year         = {2025},
  howpublished = {\url{https://attack.mitre.org/}},
  note         = {Accessed: 2025-07-26}
}

@misc{nvd,
  author       = {{National Institute of Standards and Technology (NIST)}},
  title        = {National Vulnerability Database (NVD)},
  year         = {2025},
  howpublished = {\url{https://nvd.nist.gov/}},
  note         = {Accessed: 2025-07-26}
}

@inproceedings{franco2020secbot,
  title={Secbot: a business-driven conversational agent for cybersecurity planning and management},
  author={Franco, Muriel Figueredo and Rodrigues, Bruno and Scheid, Eder John and Jacobs, Arthur and Killer, Christian and Granville, Lisandro Zambenedetti and Stiller, Burkhard},
  booktitle={2020 16th international conference on network and service management (CNSM)},
  pages={1--7},
  year={2020},
  organization={IEEE}
}

@inproceedings{rajapaksha2024rag,
  title        = {A RAG-based Question-Answering Solution for Cyber-Attack Investigation and Attribution},
  author       = {Rajapaksha, S. and Rani, R. and Karafili, E.},
  booktitle    = {European Symposium on Research in Computer Security},
  pages        = {238--256},
  year         = {2024},
  publisher    = {Springer Nature Switzerland},
  address      = {Cham},
  doi          = {https://doi.org/10.1007/978-3-031-82362-6_15},
  note         = {Published in the proceedings of the European Symposium on Research in Computer Security (ESORICS) 2024}
}

@inproceedings{arikkat2024intellbot,
  title        = {Intellbot: Retrieval Augmented LLM Chatbot for Cyber Threat Knowledge Delivery},
  author       = {Arikkat, D. R. and Abhinav, M. and Binu, N. and Parvathi, M. and Biju, N. and Arunima, K. S. and Conti, M.},
  booktitle    = {2024 IEEE 16th International Conference on Computational Intelligence and Communication Networks (CICN)},
  pages        = {644--651},
  year         = {2024},
  month        = {December},
  publisher    = {IEEE},
  doi          = {},
  note         = {Conference held at [location if known, otherwise omit]}
}

@article{zhang2026ambiguous,
  title     = {From Ambiguous Queries to Verifiable Insights: A Task-Driven Framework for LLM-Powered SOC Analysis},
  author    = {Zhang, H. and Wang, H. and Tan, H. and Zeng, L. and Li, J. and Gu, Z.},
  journal   = {CAAI Transactions on Intelligence Technology},
  year      = {2026},
  publisher = {Institution of Engineering and Technology (IET) on behalf of the Chinese Association for Artificial Intelligence (CAAI)},
  doi       = {},
  note      = {Forthcoming/Just published. DOI and page numbers not yet available.}
}

@inproceedings{paduraru2024cyberguardian,
  title        = {CyberGuardian: An Interactive Assistant for Cybersecurity Specialists Using Large Language Models},
  author       = {Ciprian Paduraru and Catalina Patilea and Alin Stefanescu},
  year         = {2024},
  month        = {July},
  booktitle    = {Proceedings of the 19th International Conference on Software Technologies (ICSOFT 2024)},
  volume       = {1},
  pages        = {442--449},
  publisher    = {SciTePress},
  address      = {Dijon, France},
  doi          = {10.5220/0012811700003753},
  isbn         = {978-989-758-706-1},
  url          = {https://www.scitepress.org/PublishedPapers/2024/128117/}
}

@inproceedings{mitra2024localintel,
  title        = {LocalIntel: Generating Organizational Threat Intelligence from Global and Local Cyber Knowledge},
  author       = {Mitra, Shaswata and Neupane, Subash and Chakraborty, Trisha and Mittal, Sudip and Piplai, Aritran and Gaur, Manas and Rahimi, Shahram},
  year         = {2024},
  month        = {December},
  booktitle    = {Foundations and Practice of Security: 17th International Symposium, FPS 2024, Montréal, QC, Canada, December 9–11, 2024, Revised Selected Papers, Part II},
  publisher    = {Springer Nature Switzerland},
  address      = {Cham},
  pages        = {63--78},
  doi          = {10.1007/978-3-031-87496-3_5},
  isbn         = {978-3-031-87496-3},
  volume       = {Part II},
  note         = {Selected papers from the 17th International Symposium on Foundations and Practice of Security (FPS 2024)},
  series       = {Lecture Notes in Computer Science}
}

@misc{ji2024sevenllm,
  title        = {SEvenLLM: Benchmarking, Eliciting, and Enhancing Abilities of Large Language Models in Cyber Threat Intelligence},
  author       = {Hangyuan Ji and Jian Yang and Liqun Chai and Chaoren Wei and Lin Yang and Yunfei Duan and Yunwang Wang and Tianwei Sun and Hanyu Guo and Tongliang Li and Changyu Ren and Zhoujun Li},
  year         = {2024},
  eprint       = {2405.03446},
  archiveprefix = {arXiv},
  primaryclass = {cs.CR},
  url          = {https://arxiv.org/abs/2405.03446},
  note         = {Version v1 submitted on 6 May 2024, v2 revised on 3 Jun 2024}
}

@inproceedings{levi2025cyberpal,
  title        = {CyberPal.{AI}: Empowering {LLMs} with Expert-Driven Cybersecurity Instructions},
  author       = {Levi, Matan and Allouche, Yair and Ohayon, Daniel and Puzanov, Anton},
  booktitle    = {Proceedings of the 39th AAAI Conference on Artificial Intelligence},
  year         = {2025},
  month        = {April},
  volume       = {39},
  pages        = {24402--24412},
  publisher    = {Association for the Advancement of Artificial Intelligence},
  doi          = {10.1609/aaai.v39i23.34618},
  url          = {https://doi.org/10.1609/aaai.v39i23.34618},
  note         = {Also available as arXiv preprint \url{https://arxiv.org/abs/2408.09304v1}}
}

@inproceedings{simoni2025morse,
  title        = {MoRSE: Bridging the Gap in Cybersecurity Expertise with Retrieval Augmented Generation},
  author       = {Marco Simoni and Andrea Saracino and Vinod P and Mauro Conti},
  year         = {2025},
  month        = {March},
  booktitle    = {Proceedings of the 40th ACM/SIGAPP Symposium on Applied Computing (SAC 2025)},
  pages        = {1213--1222},
  publisher    = {ACM},
  address      = {Catania, Italy},
  isbn         = {979-8-4007-0629-5},
  url          = {https://arxiv.org/abs/2407.15748},
  note         = {Also available as arXiv preprint \url{https://arxiv.org/abs/2407.15748}. Conference date: March 31 - April 4, 2025.}
}

@inproceedings{fang2025orion,
  title={Orion: A Multi-Agent Framework for Optimizing RAG Systems through Specialized Agent Collaboration},
  author={Fang, Xianxing and Xie, Liangru and Yang, Weibin and Zhang, Tianyi and Zhang, Ruitao and Wang, Hao and Wu, Di and Pan, Yushan},
  booktitle={Proceedings of the 16th International Conference on Internetware},
  pages={332--343},
  year={2025}
}

@article{habibzadeh2026large,
  title={Large language models for security operations centers: A comprehensive survey},
  author={Habibzadeh, Ali and Feyzi, Farid and Atani, Reza Ebrahimi},
  journal={Journal of Electrical and Computer Engineering},
  volume={2026},
  number={1},
  pages={3383674},
  year={2026},
  publisher={Wiley Online Library}
}

@inproceedings{zhao2025ontology,
  title={Ontology-aware rag for improved question-answering in cybersecurity education},
  author={Zhao, Chengshuai and Agrawal, Garima and Zhang, Fan and Kumarage, Tharindu and Tan, Zhen and Deng, Yuli and Chen, Ying-Chih and Liu, Huan},
  booktitle={2025 IEEE International Conference on Big Data (BigData)},
  pages={3161--3170},
  year={2025},
  organization={IEEE}
}

@misc{sigmahq_sigma,
  author       = {{SigmaHQ}},
  title        = {Sigma: Main Sigma Rule Repository},
  year         = {2026},
  howpublished = {\url{https://github.com/SigmaHQ/sigma}},
  note         = {Accessed: 2026-05-24}
}

@article{alam2025athenabench,
  title={AthenaBench: A Dynamic Benchmark for Evaluating LLMs in Cyber Threat Intelligence},
  author={Alam, Md Tanvirul and Bhusal, Dipkamal and Ahmad, Salman and Rastogi, Nidhi and Worth, Peter},
  journal={arXiv preprint arXiv:2511.01144},
  year={2025}
}

@inproceedings{liu2024cyberbench,
  title={Cyberbench: A multi-task benchmark for evaluating large language models in cybersecurity},
  author={Liu, Zefang and Shi, Jialei and Buford, John F},
  booktitle={AAAI 2024 Workshop on Artificial Intelligence for Cyber Security},
  year={2024}
}

@article{weerawardhena2025llama,
  title={Llama-3.1-foundationai-securityllm-8b-instruct technical report},
  author={Weerawardhena, Sajana and Kassianik, Paul and Nelson, Blaine and Saglam, Baturay and Vellore, Anu and Priyanshu, Aman and Vijay, Supriti and Aufiero, Massimo and Goldblatt, Arthur and Burch, Fraser and others},
  journal={arXiv preprint arXiv:2508.01059},
  year={2025}
}

@inproceedings{es2024ragas,
  title={Ragas: Automated evaluation of retrieval augmented generation},
  author={Es, Shahul and James, Jithin and Anke, Luis Espinosa and Schockaert, Steven},
  booktitle={Proceedings of the 18th conference of the european chapter of the association for computational linguistics: system demonstrations},
  pages={150--158},
  year={2024}
}

@article{mandilara2025decoding,
  title={Decoding the mystery: How can LLMs turn text into cypher in complex knowledge graphs?},
  author={Mandilara, Ioanna and Androna, Christina Maria and Fotopoulou, Eleni and Zafeiropoulos, Anastasios and Papavassiliou, Symeon},
  journal={IEEE Access},
  year={2025},
  publisher={IEEE}
}

@article{aghaei2025securebert,
  title={SecureBERT 2.0: Advanced language model for cybersecurity intelligence},
  author={Aghaei, Ehsan and Jain, Sarthak and Arun, Prashanth and Sambamoorthy, Arjun},
  journal={arXiv preprint arXiv:2510.00240},
  year={2025}
}


\end{document}